\documentclass[amsmath,amssymb]{appolb}
\usepackage{graphicx,mathtools}
\usepackage{hyperref}

\begin{document}
% \eqsec  % uncomment this line to get equations numbered by (sec.num)
\title{Leptogluons in dilepton production at LHC%
\thanks{Presented by D. Zhuridov at the XXXIX International Conference of Theoretical Physics 
``Matter To The Deepest", Ustro\'n, Poland, September 13-18, 2015.}%
% you can use '\\' to break lines
}
\author{Tomasz Jeli\'nski, Dmitry Zhuridov
\address{Institute of Physics, University of Silesia,\\ Uniwersytecka 4, 40-007, Katowice, Poland}
\\
%{Third Author of different affiliation
%}
%the Name(s) of other Author(s)
%\address{affiliation}
}
\maketitle
\begin{abstract}
In the composite models with colored substructure of the fermions the color singlet leptons 
are accompanied by a composite color octet partners, which are known as leptogluons. 
We consider the effect of leptogluons in the dilepton production at the LHC 
and show that in the reachable parameter range this effect is typically dominated by $t$-channel 
leptogluon exchange (indirect channel). 
We show that this channel alone can give a sizable contribution to the dimuon production 
at the LHC for TeV scale values of the invariant mass of $\mu^+\mu^-$ pairs.
\end{abstract}

\section{Introduction}
For about a century particle physics investigates matter at  distances from about $10^{-10}$ m (size of the atom) 
to about $10^{-15}$ m (nucleons substructure), so it is five orders of magnitude progress in exploring micro-world. 
A big question is what can happen next? Do presently known as elementary particles are complex at smaller distances? There are many interesting theories which 
explore physics at these tiny distances below $10^{-15}$ m, let us mention only theories of extra dimensions or string theories.
Yet another type of models constitute so-called composite 
models~\cite{Pati:1974yy,Terazawa:1976xx,Shupe:1979fv,Harari:1979gi,Squires:1980cm,Harari:1980ez,Barbieri:1980aq,Fritzsch:1981zh}.

Early models, which introduced a substructure of the Standard Model (SM) leptons, were discussed 
in Refs.~\cite{Harari:1979gi,Squires:1980cm,Harari:1980ez,Barbieri:1980aq,Fritzsch:1981zh}.  
Leptons with %(at least two) 
colored subcomponents are automatically accompanied 
by a color octet composites $\ell_8$ having the same lepton numbers, which are called {\it leptogluons}. %~\cite{Schrempp:1988ku},
%belong to the adjoint representation of the $SU(3)$ group, and 
They can be probed at the high-energy collider experiments~\cite{Abolins:1982vy,Hewett:1997ce,AbelleiraFernandez:2012cc}, 
in particular, at the LHC frontier~\cite{Celikel:1998dj,Mandal:2012rx,Goncalves-Netto:2013nla}. 
Collider effects of the leptogluons are of exceptional interest since they are dominated by the tree level processes, while 
the related contact interactions and contributions to the lepton magnetic moments 
have one- and two-loop suppression, respectively.

The strongest mass bound for the charged leptogluons is $m_8>1.2$-$1.3$~TeV~\cite{Goncalves-Netto:2013nla}.
However for the choice of parameters in Ref.~\cite{Goncalves-Netto:2013nla} the $t$-channel production of leptogluons 
is suppressed with respect to their pair production. In this work we show that for the 
compositeness scale $\Lambda$, 
which is close to the allowed values of $m_8$, the $t$-channel exchange of leptogluons %the indirect production channel of leptogluons 
dominates over their pair production at 8~TeV LHC, and this channel alone can give 
a sizable contribution to the production of dileptons with the invariant mass $m(\ell^+\ell^-)=\mathcal{O}(1)$~TeV. % at the LHC. 
(Here and below $m_8$ denotes the relevant $\ell_8^\pm$ mass).

%%%%%%%%%%%%%%%%%%%%%%%%%%%%%%%%%%%%%%%%%%%%%%%%%%%%%%%%%%%%%%%%%%%%%%%%%%%%%%%%%%%%%%%%%%%%%%%%%%%%%%%%%%%%%
\section{Indirect and pair production of leptogluons at the LHC}\label{section:LHC}
Effective interaction of $\ell_8$ with leptons and gluons can be 
written as\footnote{Notice that the effective compositeness scale for 
contact (4-fermion) interactions %$\psi\psi\psi\psi$ 
may exceed the scale $\Lambda$ in Eq.~(\ref{eq:Lagr})  
due to the loop factor, which was mentioned above. Notice that factor 1/2 in Eq.~(\ref{eq:Lagr}) 
leads to the Feynman rule without factor 2.}~\cite{PDG2014}
\begin{eqnarray}\label{eq:Lagr}
		\mathcal{L} = \frac{g_s}{2\Lambda} 
		\overline{\ell_8^A}\sigma^{\mu\nu}G_{\mu\nu}^A(a_{\ell L}P_L+a_{\ell R}P_R)\ell + {\rm H.c.},
\end{eqnarray}
where $g_s$ is the strong coupling constant, $G_{\mu\nu}^A$ 
is the gluon field strength, $P_{L(R)}$ is the left (right) projector, $\ell=e,\mu,\tau$, 
$\sigma^{\mu\nu}=\frac{i}{2}[\gamma^\mu,\gamma^\nu]$, and for the new couplings we take: 
$a_{\ell L}=1$ and $a_{\ell R}=0$~\cite{Goncalves-Netto:2013nla}. 
The width of the dominant decay of $\ell_8$ can be written as 
$\Gamma_{\ell_8\to g\ell} =  \alpha_s m_8^{3}/(4\Lambda^2)$,
%\footnote{Notice that $\Gamma_{\ell_8\to g\ell}$ %in Eq.~(\ref{eq:width}) 
%depends on the definition of $\mathcal{L}$ in Eq.~(\ref{eq:Lagr}). 
%Sometimes factor $1/2$ is omitted  in Eq.~(\ref{eq:Lagr})~\cite{Hewett:1997ce}. %,BowserChao:1998mh}. 
%Then there is no factor $1/4$ in the width.}
%\begin{eqnarray}\label{eq:width}
%		\Gamma_{\ell_8\to g\ell} =  \alpha_s (a_{\ell L}^2+a_{\ell R}^2)  \frac{m_8^{3}}{4\Lambda^2},
%\end{eqnarray}
where $\alpha_s=g_s^2/(4\pi)$. 
Below we consider long-lived leptogluons with $\Gamma\ll m_8$. % with the mass $m_8$.

\vspace{-2mm}
%%%%%%%%%%%%%%%%%%%%%%%%%%%%%%%%%%%%%%%%%%
 \begin{figure}[htb]
 \centerline{%
 	\includegraphics[width=0.45\textwidth]{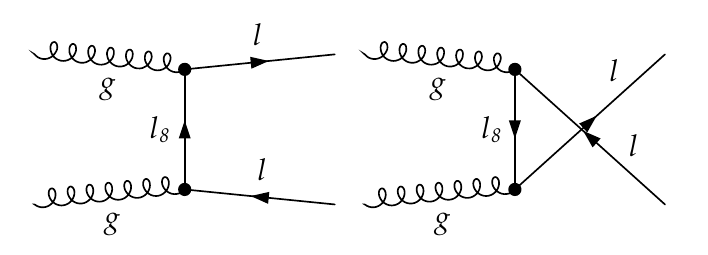} }
	   \caption{Leading Feynman diagrams for $gg\to\ell^+\ell^-$ via $t$-channel exchange of $\ell_8^\pm$.}
   \label{Fig:diagrams_indir}
 \end{figure}
%%%%%%%%%%%%%%%%%%%%%%%%%%%%%%%%%%%%%%%%%%
\vspace{-2mm}
 \begin{figure}[htb]
 \centerline{%	
 	\includegraphics[width=0.52\textwidth]{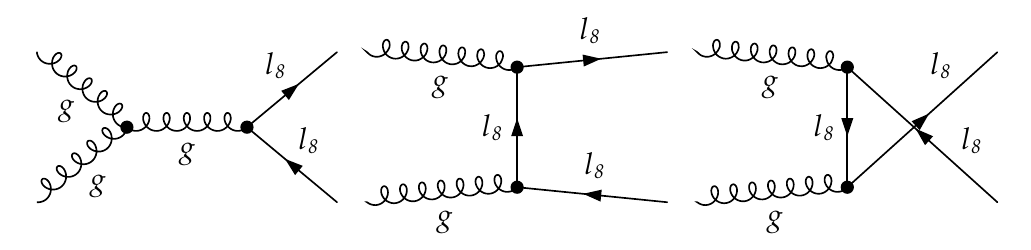}\hspace{-2mm}
 	\includegraphics[width=0.32\textwidth]{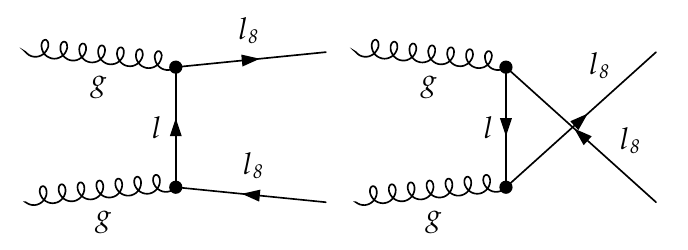}\hspace{-2mm}
 	\includegraphics[width=0.19\textwidth]{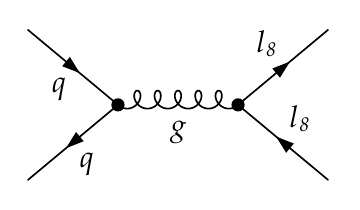}}
	   \caption{Leading Feynman diagrams for the processes $gg\to\ell_8^+\ell_8^-$ and $q\bar q\to\ell_8^+\ell_8^-$.}
   \label{Fig:diagrams_pair}
 \end{figure}
%%%%%%%%%%%%%%%%%%%%%%%%%%%%%%%%%%%%%%%%%%
The leading Feynman diagrams on the parton level for {\it indirect production} (IP) and 
{\it pair production} (PP)\footnote{Directly produced 
$\ell_8^\pm$ %undergo a subsequent radiative 
undergo $\ell_8^\pm\to\ell^\pm g$ decays with close to 100\% branching ratios.} 
of $\ell_8$ in $p$-$p$ collisions are shown 
in Figs.~\ref{Fig:diagrams_indir} and \ref{Fig:diagrams_pair}, respectively, 
and the total cross sections are %for the respective subprocesses can be written as 
(see Appendix~\ref{app:A}) % and ~\ref{app:B})

\begin{eqnarray}\label{eq:section_gg-mumu}
	\hat\sigma_{gg\to \ell^+\ell^-} &=& \frac{\pi}{12}\alpha_s^2\,\xi^4\, m_8^2\  
	    F\left( r \right), \\
	\hat\sigma_{q\bar q\to \ell_8^+\ell_8^-} &=& \frac{16\pi}{9}\frac{\alpha_s^2}{m_8^2}\, r(1+2r)\beta,\label{eq:qq_mu8mu8} \\
	\hat\sigma_{g g   \to \ell_8^+\ell_8^-} &=& \frac{\pi}{12}\frac{\alpha_s^2}{m_8^2} \left[ 
	    F_1(r) + \xi^4 m_8^4\, F_2(r) + \xi^2m_8^2\, F_{12}(r)  \right],   \label{eq:sigma_gg-mu8mu8}
\end{eqnarray}
where we neglected the terms of $\mathcal{O}(\Gamma/m_8)$ which effect is below 1\%, %$m_{\ell\ell}=\sqrt{\hat s}$ is the dimuon invariant mass, 
$\xi=a_{\ell L}/\Lambda$, $r=m_8^2/\hat s$, $\beta=\sqrt{1-4r}$, and other functions are defined as %$\rho=m_{\mu\mu}^2/m_8^2$ with 
\begin{eqnarray}
		F(r) &=&  \frac{1-6\,r-24\,r^2}{2\, r} +   
		3\,r(3+4r)\, {\rm ln}\left(\frac{1+r}{r}\right), \\
		F_1(r) &=&  -18\,r(4+17\,r)\beta + 54\,r(1+4\,r-4\,r^2)\,{\rm ln}\left(\frac{1+\beta}{1-\beta}\right), \\
		F_2(r) &=&  \frac{4\,(1-4\,r)}{r}  \left[ (1+6\,r)\beta + 6\,r^2\, 
		      {\rm ln}\left(\frac{1-2\,r+\beta}{1-2\,r-\beta}\right)  \right], \\
		F_{12}(r) &=&  -3\,(2+r)(1+6\,r)\beta \nonumber\\
		       &+& \frac{18\,r(1+r)}{1-r}\left[ 
		      {\rm ln}\left(\frac{1+\beta}{1-\beta}\right) 
		      + r^2\, {\rm ln}\left(\frac{1-2\,r+\beta}{1-2\,r-\beta}\right) \right].  \label{eq:F12}
\end{eqnarray}

The total cross section for ${pp\to abX \to cdX}$ can be calculated as 
\begin{eqnarray}
	\sigma_{pp\to cd X} &=&	\int\limits_{y_0}^1 \frac{dy}{y} \int\limits_{y}^1 \frac{dx}{x}\, p_a(x,\mu_F^2) 
	\, p_b\left(\frac{y}{x},  \mu_F^2\right)  \hat\sigma_{ab\to cd}(ys),
	%\nonumber\\
	%&=&  \int\limits_{y_0}^{1} \frac{dy}{y} \, S_{pp\to cd X}(ys),
\end{eqnarray}
where $y_0=\mu_{cd}^2/s$ ($\mu_{cd}$ is the minimal invariant mass of $cd$), $\sqrt{s}$ is the total energy of 
the proton-proton collisions, $\mu_F$ is the factorization scale, %(we use $\mu_F=m_8$~\cite{Goncalves-Netto:2013nla}), 
$p_a(x,Q^2)=x\,{\rm pdf}_a(x,Q)$ 
%\begin{eqnarray}
%	p_a(x,Q^2)=x\, {\rm pdf}_a(x,Q)
%\end{eqnarray}
is the parton $a$ distribution in proton for the momentum transfer $Q$, and $X$ represents the two jets close to 
the beam axis. 
%
%Notice that the differential cross sections given in the Appendix~\ref{app:A} %and ~\ref{app:B} 
%may help in determination of the properties of leptogluons. 
%%may help in distinguishing leptogluons from leptoquarks~\cite{CMS:2014qpa}, etc.

\begin{figure}[htb]
 \centerline{%
 	\includegraphics[width=0.52\textwidth]{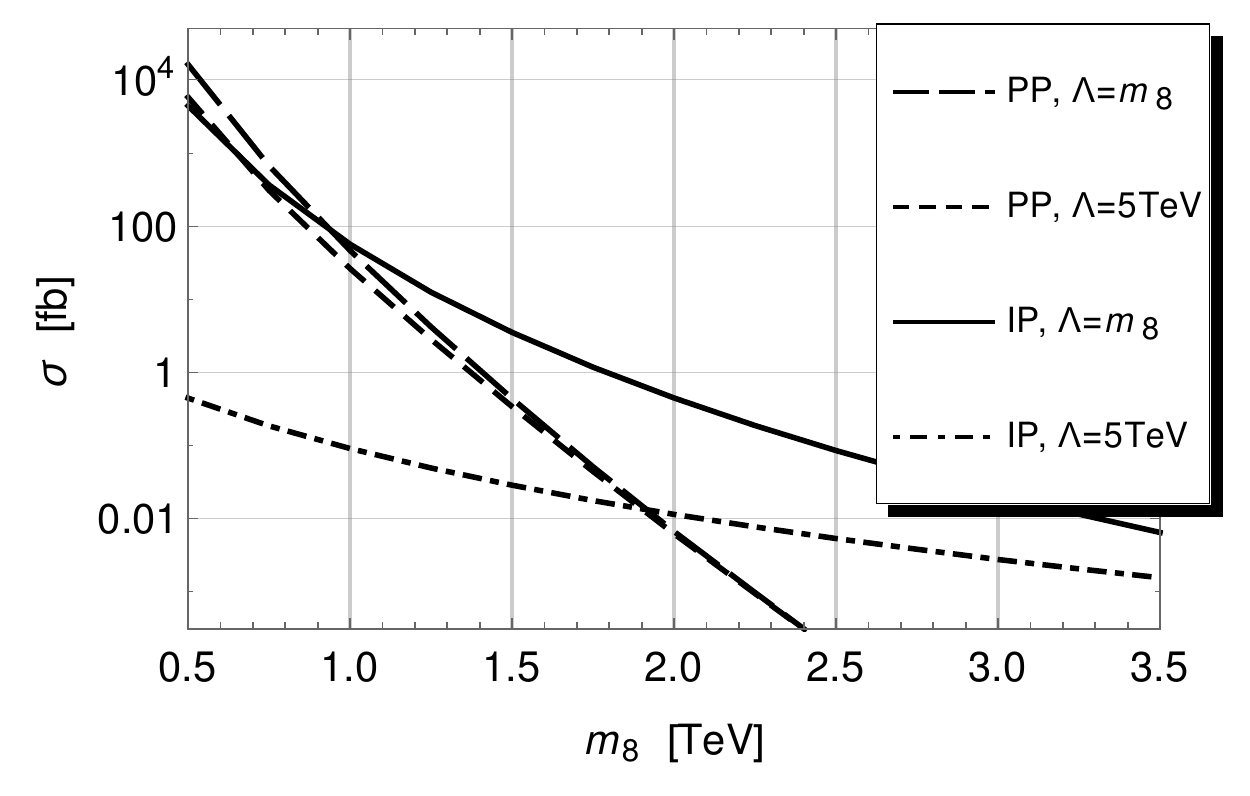}
 	\includegraphics[width=0.48\textwidth]{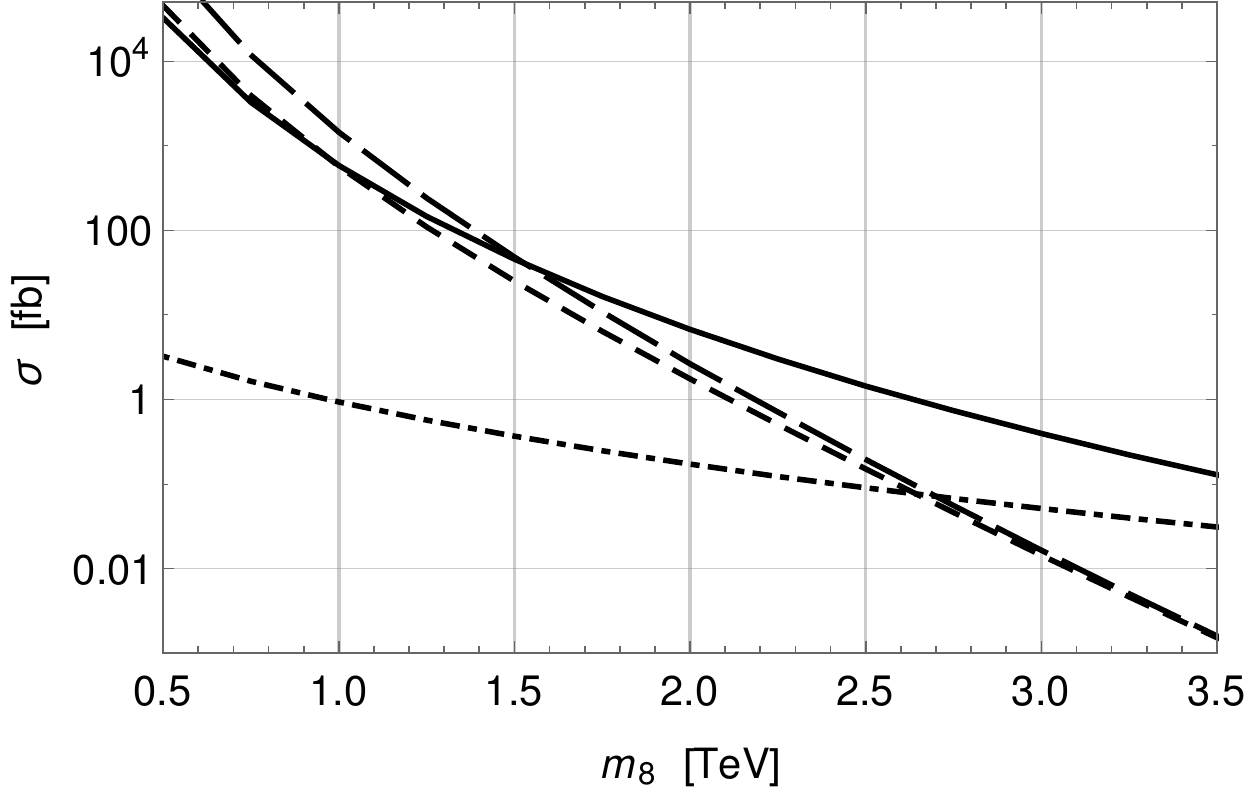}}
	   \caption{Total cross sections for various processes that involve leptogluons versus the 
	   leptogluon mass $m_8$ for $\sqrt{s}=8$~TeV ({\it left}) and 14~TeV ({\it right}). 
	   Solid (dot-dashed) and long-dashed (short-dashed) lines represent 
	   $pp\xrightarrow{\ell_8}\ell^+\ell^-$ %with $t$-channel $\ell_8^\pm$ exchange (IP) 
	   and $pp\to\ell_8^+\ell_8^-$ processes 
	   for the compositeness scale $\Lambda=m_8$ ($\Lambda=5$~TeV), respectively.}
   \label{Fig:channels}
 \end{figure}

Numerical calculations we performed %using Eqs.~(\ref{eq:section_gg-mumu})-(\ref{eq:F12})  
in MadGraph5~\cite{Alwall:2011uj}, using FeynRules~\cite{Christensen:2008py,Alloul:2013bka} 
to generate 
%Universal FeynRules Output 
UFO-format~\cite{Degrande:2011ua} model files.
%and we made numerical checks 
%in Mathematica %~\cite{Math_10-2} 
%with the CTEQ5 parton distributions~\cite{Lai:1999wy}.
%
%The results of these two methods are consistent with each other within 10\% discrepancy, %in the interested energy range. 
%which is strongly affected by $\mu_F$ (we use $\mu_F=m_8$)~\cite{Goncalves-Netto:2013nla}.
Fig.~\ref{Fig:channels} shows cross sections for IP and PP\footnote{The dependence of PP of $\ell_8$ on $\Lambda$ is due to 
the 4th and 5th diagrams in Fig.~\ref{Fig:diagrams_pair}.} of leptogluons at the LHC. In particular, 
IP of $\ell_8$ dominates at 8\,TeV LHC for $m_8>1.2$\,TeV (current bound) and $\Lambda\sim m_8$. 
For $m_8\approx1$~TeV the cross sections increase by factor of $\mathcal{O}(10)$ with the energy increase up to 14\,TeV. 
For $m_8\approx2$~TeV the PP (IP) cross section increases by factor of about 300 ($\sim30$)  
with the same energy increase. 
%which makes possible to observe $\ell_8$ at the new run of the LHC. 

%%%%%%%%%%%%%%%%%%%%%%%%%%%%%%%%%%%%%%%%%%
%%%%%%%%%%%%%%%%%%%%%%%%%%%%%%%%%%%%%%%%%% 
%%%%%%%%%%%%%%%%%%%%%%%%%%%%%%%%%%%%%%%%%%
 \begin{figure}[htb]
 \centerline{%
 	\includegraphics[width=0.48\textwidth]{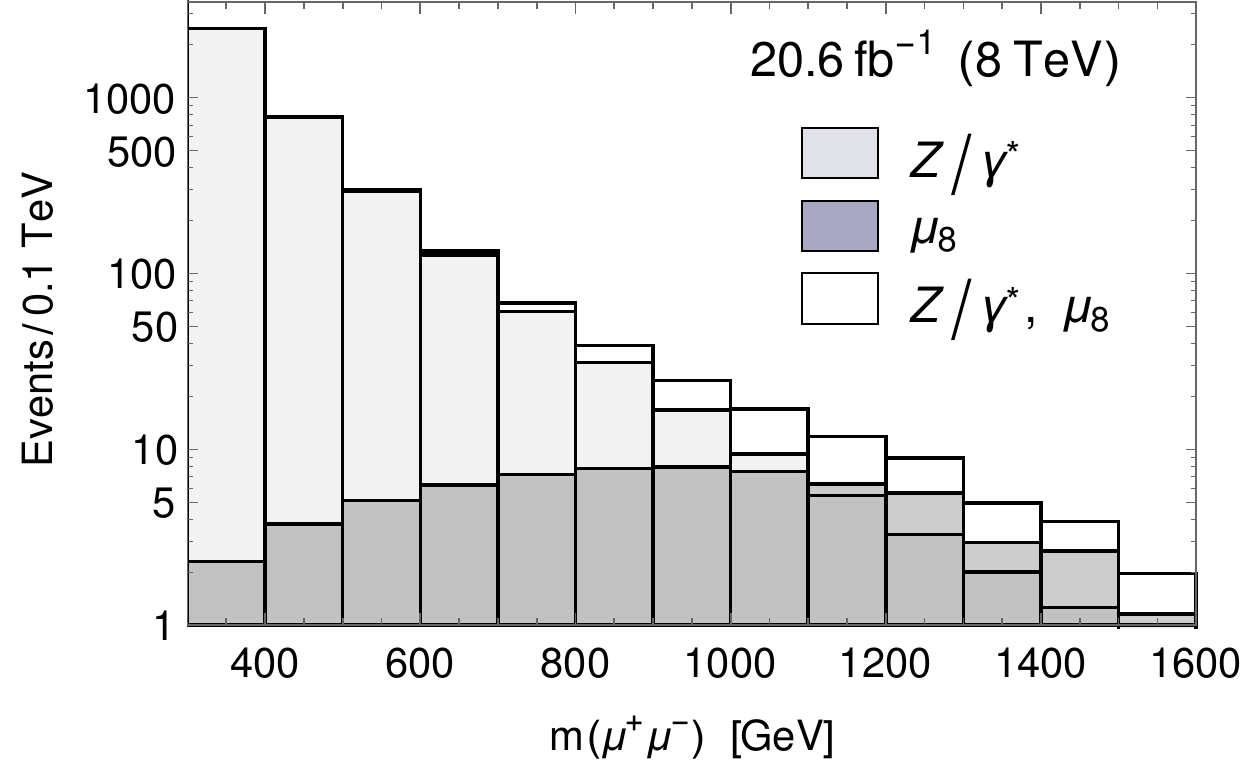} \quad
 	\includegraphics[width=0.46\textwidth]{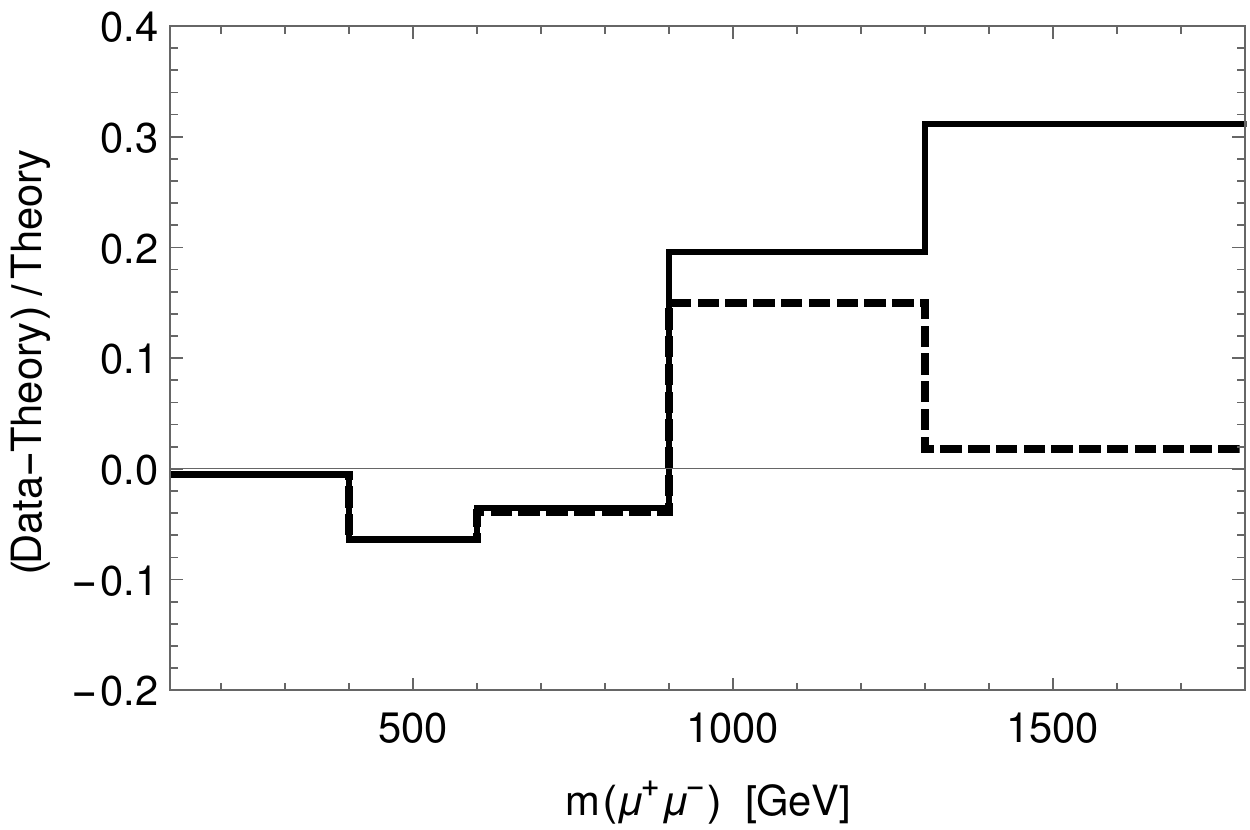}}
	   \caption{{\it Left}: Simulated $\mu^+\mu^-$ invariant mass spectra. 
	   {\it Right}: Normalized difference between the number of the CMS data and simulated dimuon 
	   events in the given $m(\mu^+\mu^-)$ ranges for 
	   $\sqrt{s}=8$~TeV and with $20.6~{\rm fb}^{-1}$.    
	   Solid (dashed) line is connected with the SM background (the SM background plus the signal of $\mu_8^\pm$).} % with $m_8=2$~TeV).}
   \label{Fig:data-bkg}
 \end{figure}
%%%%%%%%%%%%%%%%%%%%%%%%%%%%%%%%%%%%%%%%%% 
Fig.~\ref{Fig:data-bkg} (left) shows the simulated dimuon invariant mass spectra at the LHC with 
$\sqrt{s}=8$~TeV and $20.6$~fb$^{-1}$ of integrated luminosity, where light, 
dark and white histograms represent Drell-Yan production (dominant SM background: $Z/\gamma^*$), 
the effect of muonic leptogluons 
$\mu_8^\pm$ with $m_8=\Lambda=1.5$ TeV, and their combination, recpectively. 
The difference between the number of the CMS data~\cite{Khachatryan:2014fba} and simulated events normalized to the simulated 
events in various ranges of the invariant mass $m(\mu^+\mu^-)$ is shown in Fig.~\ref{Fig:data-bkg} (right). 
The solid line is connected with the SM background. 
The dashed line corresponds to the combination of the SM background and the effect of IP of $\mu_8^\pm$ 
with the mass $m_8=2$~TeV and coupling-to-scale ratio $\xi=(2.4~{\rm TeV})^{-1}$, 
which minimizes the likelihood function: $\chi^2_{{\rm min}}=2.07$. 
Fig.~\ref{Fig:data-bkg} shows that IP of $\mu_8$ decreases the dimuon signal 
for large $m(\mu^+\mu^-)$.
%may essentially reduce the
%deviations of the central values of the CMS data points from the SM 
%background in the region of large $m(\mu^+\mu^-)$. 
%%Although the current ``deviations'' are within $1\sigma$ errorbars this result is useful since 
%%the LHC2 data will have essentially smaller errorbars. 

To conclude, the present analysis shows a possibility of sizable effects of leptogluons 
in dilepton production at the LHC for large invariant masses.
%More data will help to further constrain the parameters, and the introduced analytical expressions 
%can be used to discriminate between various models with the leptogluons, leptoquarks, etc.

%\vspace{-5mm}
%%%%%%%%%%%%%%%%%%%%%%%%%%%%%%%%%%%%%%%%%%%%%%%
\section*{Acknowledgements} 

We would like to thank Janusz Gluza and Henryk Czy\.z for collaborative work. 
%, and the participants of the 
%XXXIX International Conference of Theoretical Physics ``Matter To The Deepest" for useful discussions and comments. 
This work was supported in part by the Polish National Science Centre, grant number 
DEC-2012/07/B/ST2/03867.

%\vspace{-5mm}
%%%%%%%%%%%%%%%%%%%%%%%%%%%%%%%%%%%%%%%%%%%%%%%
\appendix
\section{}\label{app:A}

\subsection{\quad Indirect production of $\ell_8^\pm$} 

Analytical results were derived with the help of \mbox{FeynArts}~\cite{Hahn:2000kx} and FormCalc~\cite{Hahn:1998yk}. 
Differential cross section for %$gg\to\ell^+\ell^-$ via $t$-channel $\ell_8^\pm$ exchange 
IP of leptogluons %$\ell_8^\pm$ 
can be written as
\begin{eqnarray}\label{eq:diff_cross-section}
		\frac{d\hat\sigma_{gg\to \ell^+\ell^-}}{d\hat t} = \frac{1}{16\pi\hat s^2}\frac{1}{256}\, d_R\, g_s^4\xi^4 
		\sum(\mathcal{M}_{11} + \mathcal{M}_{22} ),
\end{eqnarray}
where the two summands (one of them is missing in Ref.~\cite{Akay:2010sw}) 
correspond to the two diagrams in Fig.~\ref{Fig:diagrams_indir}, %(interference is zero), 
$d_R=8$ is the dimension of octet representation of $SU(3)$, 
factor $1/256=1/(2^2\,8^2)$ 
comes from the averaging over %the degrees of freedom 
polarizations and colors of gluons, and 
normalized squared matrix elements are %given as follows
\begin{eqnarray}\label{eq:M_11}
		\sum\mathcal{M}_{11}  = -\frac{4\,\hat t^3 (\hat s+\hat t)}
		{(\hat t-m_8^2)^2},	\quad  
		\sum\mathcal{M}_{22}  = -\frac{4\,\hat t 
		(\hat s+\hat t)^3}{(\hat u-m_8^2)^2}, \label{eq:M_22}
\end{eqnarray}
where $\hat s=(k_1+k_2)^2$, $\hat t=(q_1-k_1)^2$ and $\hat u=(q_2-k_1)^2$ are the Mandelstam variables, 
%for the considered subprocess. 
and $\sum$ denotes the summation over initial and final spin states. 
Then Eq.~(\ref{eq:section_gg-mumu}) can be derived using the formula
\begin{eqnarray}%\label{eq:vertex}
		\hat\sigma_{gg\to \ell^+\ell^-} = \int\limits_{-\hat s}^0 
		d\hat t\ \frac{d\hat\sigma_{gg\to \ell^+\ell^-}}{d\hat t}.
\end{eqnarray}

%%%%%%%%%%%%%%%%%%%%%%%%%%%%%%%%%%%%%%%%%%%%%%%
\subsection{\quad $\ell_8^+\ell_8^-$ pair production}%\label{app:B}

%%%%%%%%%%%%%%%%%%%%%%%%%%%%%%%%%%%%%%%%%%%%%%%%
%\subsection{Process \boldmath{$gg\to\ell_8^+\ell_8^-$}}
%
Following the method of Refs.~\cite{Combridge:1978kx,Georgi:1978kx,Tanaka:1991nr}  
for $gg\to\ell_8^+\ell_8^-$ we have
\begin{eqnarray}\label{eq:diff_section}
		\frac{d\hat\sigma_{gg\to \ell_8^+\ell_8^-}}{d\hat t} &=& \frac{\pi\alpha_s^2}{16\hat s^2}
		    \left[ K_1(R)\sum(\mathcal{M}_{ss}+\mathcal{M}_{st}
		    +\mathcal{M}_{su})\right. \nonumber\\ 
		    &+& K_2(R)\sum(\mathcal{M}_{tt}+\mathcal{M}_{uu}) + K_3(R)\sum\mathcal{M}_{tu} \nonumber\\
		    &+& \xi^4\, K_4(R)\sum(\mathcal{M}_{tt}^{\ell\ell}+\mathcal{M}_{uu}^{\ell\ell}) \nonumber\\
		    &+& \left.\xi^2\, K_5(R)\sum(\mathcal{M}_{st}^\ell+\mathcal{M}_{su}^\ell)  
		    %&+&  \xi^2\, K_6(R)\sum(\mathcal{M}_{tt}^\ell-\mathcal{M}_{uu}^\ell) 
			   + \xi^2\, K_6(R)\sum\mathcal{M}_{tu}^\ell  \right],
\end{eqnarray}
where the terms with $\mathcal{M}_{tt}^{\ell}$ and $\mathcal{M}_{uu}^{\ell}$ are absent due to zero color factors, 
and the normalized squared matrix elements are given as follows
\begin{eqnarray}%\label{eq:vertex}
		\sum\mathcal{M}_{ss}  &=& \, \frac{(\hat t - m^2)(\hat u - m^2)}{\hat s^2}, \\
		\sum\mathcal{M}_{st}  &=& \, \frac{(\hat t - m^2)(\hat u - m^2) 
		      + m^2(\hat u - \hat t)}{2\hat s (\hat t - m^2)} = \sum\mathcal{M}_{su}
		      (\hat t\leftrightarrow\hat u), \label{eq:interf_st}	\\
		%\sum\mathcal{M}_{su}  &=& \, \frac{(\hat t - m^2)(\hat u - m^2) 
		%      + m^2(\hat t - \hat u)}{2\hat s (\hat u - m^2)}, \label{eq:interf_su} \\
		\sum\mathcal{M}_{tt}  &=& \frac{(\hat t - m^2)(\hat u - m^2) 
		      - 2m^2(\hat t + m^2)}{2(\hat t - m^2)^2} = \sum\mathcal{M}_{uu}(\hat t\leftrightarrow\hat u),	\\
		%\sum\mathcal{M}_{uu}  &=& \frac{(\hat t - m^2)(\hat u - m^2) 
		%      - 2m^2(\hat u + m^2)}{2(\hat u - m^2)^2},  \\
		\sum\mathcal{M}_{tu}  &=& -\, \frac{m^2(\hat s - 4m^2)}{2(\hat t - m^2)(\hat u - m^2)}, \\
%\end{eqnarray}
%\begin{eqnarray}
		\sum\mathcal{M}_{tt}^{\ell\ell}  &=& \frac{(\hat t\hat u - m^4)(\hat t - m^2)^2}{4\hat t^2}  =  
			\sum\mathcal{M}_{uu}^{\ell\ell}(\hat t\leftrightarrow\hat u),	\\
		%\sum\mathcal{M}_{uu}^{\ell\ell}  &=& \frac{(\hat t\hat u - m^4)(\hat u - m^2)^2}{4\hat u^2},  \\
		\sum\mathcal{M}_{st}^\ell  &=& \frac{\hat t\hat u -4\hat t^2 +\hat u^2 + m^2(13\hat t -\hat u)}{8\hat s} 
			- m^4\, \frac{8\hat t +\hat u -4m^2}{4\hat s\hat t}   \nonumber  \\
			&-&  5\,m^2\, \frac{\hat t -m^2}{8\hat t}  =  \sum\mathcal{M}_{su}^\ell(\hat t\leftrightarrow\hat u), \\
		%\sum\mathcal{M}_{su}^\ell  &=& \frac{\hat t\hat u +\hat t^2 -4\hat u^2 + m^2(-\hat t +13\hat u)}{8\hat s} 
		%	- m^4\, \frac{\hat t +8\hat u -4m^2}{4\hat s\hat u}   \nonumber  \\
		%	&-&  5\,m^2\, \frac{\hat u -m^2}{8\hat u}, 
%\end{eqnarray}
%\vspace{-7mm}
%\begin{eqnarray}
		\sum\mathcal{M}_{tu}^\ell  &=& \left[  -\,\frac{(\hat t\hat u - m^4)(\hat t + 2m^2)}{8(\hat t - m^2)\hat u} \right.\nonumber\\
		  &+&  \left. m^2\frac{\hat t\hat u -4\hat u^2 +2m^2(3\hat t+7\hat u)-17m^4}{8(\hat t - m^2)\hat u} \right] 
		   +   [\hat t\leftrightarrow \hat u],
		%			   -\,\frac{(\hat t\hat u - m^4)(\hat u + 2m^2)}{8(\hat u - m^2)\hat t} + 
		%	m^2\frac{\hat t\hat u -4\hat t^2 +2m^2(7\hat t+3\hat u)-17m^4}{8(\hat u - m^2)\hat t},
\end{eqnarray}
where $m\equiv m_8$, and the nonvanishing color factors can be written as
\begin{eqnarray}%\label{eq:vertex}
		&& K_1(R) =   %f_{ABE}f_{ABF}\, \text{tr}( t_Et_F ) = 2\, i f_{ABE}\, \text{tr}( t_Et_Bt_A )  = 
	d_R\,C_A\,C_F =72, \quad
		K_2(R) =   %\text{tr}( t_At_Bt_Bt_A )   = 
	d_R\,C_F^2 = 72, \\
		&& K_3(R) =   %-2\,\text{tr}( t_At_Bt_At_B )   = 
	d_R\,C_F[C_A-2C_F] = -72, \\
		&& K_4(R) = %\delta_{AA}\delta_{BB}  =  
	64, \quad
		K_5(R) =  %i\,f_{ABE}(t_E)_{AB}   =  f_{ABE}f_{EAB}  =  
	- K_6(R) = 24,%\\
	%	K_6(R) &=&  %(t_At_B)_{BA} = - f_{ABE}f_{ABE} = 
	%-24,
\end{eqnarray}
where %$d_R$ is the dimension of the octet representation $R$ of $SU(3)$ group, and 
$C_A$ %(with $N=3$ the dimension of the group) 
and $C_F$ are the Casimir invariants.
In our case of $SU(3)$ octets we have $d_R=8$ and $C_A=C_F=3$. 
Eq.~(\ref{eq:sigma_gg-mu8mu8}) can be derived using the formula
\begin{eqnarray}\label{eq:intergation}
		\hat\sigma_{gg\to \ell_8^+\ell_8^-} = \int\limits_{m^2-\frac{\hat s}{2}(1+\beta)}^
		{m^2-\frac{\hat s}{2}(1-\beta)} d\hat t\ \frac{d\hat\sigma_{gg\to \ell_8^+\ell_8^-}}{d\hat t}.
\end{eqnarray}
%%where $\beta=\sqrt{1-4m^2/\hat s}$, 

The terms that include $\xi$ in Eq.~(\ref{eq:diff_section}) are new analytical results 
related to the 4th and 5th diagrams in Fig.~\ref{Fig:diagrams_pair} and their interference with others.
%first two lines of Eq.~(\ref{eq:diff_section}), which correspond to the set of the first three diagrams 
%in Fig.~\ref{Fig:diagrams_pair}, agree with Refs.~\cite{Combridge:1978kx,Tanaka:1991nr}. 

The differential cross section for $q\bar q\to\ell_8^+\ell_8^-$ is given in Ref.~\cite{Celikel:1998dj}. 
However there is a misptint in Ref.~\cite{Celikel:1998dj} %,Errede:1984mc}
concerning the interference terms in Eq.~(\ref{eq:interf_st}).

%%%%%%%%%%%%%%%%%%%%%%%%%%%%%%%%%%%%%%%%%%%%%%%%%%%%%%%%%%%%%%%%%%%%%%%%%%%%%%%%%%%%%%%%%%%%%%%%%%%%%%%%%%%%%

\end{document}